\documentclass[journal,10pt,twocolumn]{IEEEtran}

\usepackage{epsfig}
\usepackage{amsfonts,amsmath,amssymb,amstext}
\usepackage{times,wasysym}
\usepackage{graphicx}
\usepackage{dsfont}
\usepackage{color}
\usepackage{cite}

\title{Multi-layer Space Information Networks: Access Design and Softwarization}
\author{Hayder Al-Hraishawi,~\IEEEmembership{Member,~IEEE},
Mario Minardi,~\IEEEmembership{Student Member,~IEEE},\\
Houcine Chougrani,~\IEEEmembership{Member,~IEEE}, Oltjon Kodheli,~\IEEEmembership{Student Member,~IEEE}, \\ Jesus Fabian Mendoza Montoya, and Symeon Chatzinotas,~\IEEEmembership{Senior Member,~IEEE} 
\thanks{
The authors are with the Interdisciplinary Centre for Security, Reliability and Trust (SnT), University of Luxembourg, Luxembourg. \textit{Corresponding author: Hayder Al-Hraishawi (hayder al-hraishawi@uni.lu).}
}
\thanks{This research was funded in whole by the Luxembourg National Research Fund (FNR) in the frameworks of the FNR-CORE project "MegaLEO: Self-Organised Lower Earth Orbit Mega-Constellations"  (Grant no. C20/IS/14767486). For the purpose of open access, the authors have applied a Creative Commons Attribution 4.0 International (CC BY 4.0) license to any Author Accepted Manuscript version arising from this submission.}}

\begin{document}
\maketitle

\begin{abstract}
In this paper, we propose an approach for constructing a multi-layer multi-orbit space information network (SIN) to provide high-speed continuous broadband connectivity for space missions (nanosatellite terminals)  from the emerging space-based Internet providers. 
This notion has been motivated by the rapid developments in satellite technologies in terms of satellite miniaturization and reusable rocket launch, as well as the increased number of nanosatellite constellations in lower orbits for space downstream applications, such as earth observation, remote sensing, and Internet of Things (IoT) data collection. 
Specifically, space-based Internet providers, such as Starlink, OneWeb, and SES O3b, can be utilized for broadband connectivity directly to/from the nanosatellites, which allows a larger degree of connectivity in space network topologies. Besides, this kind of establishment is more economically efficient and eliminates the need for an excessive number of ground stations while achieving real-time and reliable space communications. This objective necessitates developing suitable radio access schemes and efficient scalable space backhauling using inter-satellite links (ISLs) and inter-orbit links (IOLs). Particularly, service-oriented radio access methods in addition to software-defined networking (SDN)-based architecture employing optimal routing mechanisms over multiple ISLs and IOLs are the most essential enablers for this novel concept.
Thus, developing this symbiotic interaction between versatile satellite nodes across different orbits will lead to a breakthrough in the way that future downstream space missions and satellite networks are designed and operated.
\end{abstract}


\section{Introduction}
The recent swift developments in satellite technologies, in particular satellite and component miniaturization and reusable rocket carrying multiple small satellites in one launch, have facilitated space accessibility through relatively inexpensive means \cite{Babich2020}. As space technologies are becoming cheaper, closer, and smaller, space industry is reviving and offering various applications spanning a wide range of services, e.g., earth observation, space observation, spectrum monitoring, asset tracking, remote sensing, space-based cloud, and Internet of Things (IoT) data collection \cite{NGSO_survey}. 
Moreover, developing satellite communication infrastructure systems in conjunction with the evolving B5G/6G networks has been a trending topic in both academia and industry. For instance, the standard development organizations, including the 3rd Generation Partnership Project (3GPP), the International Telecommunication Union (ITU), and the European Telecommunications Standards Institute (ETSI), have been investigating the use of satellite communication networks to integrate space and terrestrial communication networks in order to support future wireless ecosystems \cite{3GPP38821v16}.

 In light of the recent satellite technological progress, constructing and launching a chain of small satellites that have short lifespans become within the bounds of practicality. This will make satellite infrastructure to be more frequently upgraded and will increase the satellite number in lower orbits for various space downstream applications \cite{Hayder_NGSO}. Along with the opportunities that are foreseen in this transformation, several vigorous communication challenges require innovation to downlink the collected data back to Earth, provide higher throughput and flexibility, and network orchestration between ground and space \cite{Mayorga2021}.
 Currently, downstream mission operators heavily depend on a network of ground stations distributed across the globe for downlinking the small satellites or controlling them through telemetry and telecommand (TT\&C), which will require a massive network of ground stations \cite{Portillo2019}.  Recently, some innovative concepts towards ground network sharing have been proposed, such as Amazon AWS ground station \cite{AWS} and Microsoft Azure Orbital \cite{Azure}, but the number and duration of ground access sessions are most of the times limited, and thus, preventing real-time mission operation and continuous high-throughput downstreaming data. Therefore, novel downlinking techniques to fulfill the increased complexity of application requirements need to be developed for such dynamic environments. 

In parallel, the notion of using large constellations of non-geostationary orbit (NGSO) satellites to provide broadband connectivity from space has gained popularity and experienced a tremendous growth in the last few years \cite{Portillo2019}. 
This trend can be seen in the impressive achievements of the advent space-based Internet providers, in particular, SpaceX Starlink, SES O3B, and OneWeb, which is reflected in the growing technological and business thrusts. 
In fact, the new proposals for large NGSO constellations are surging to provide global broadband services from space \cite{Portillo2019}. 
Specifically, based on the recent satellite database released by the Union of Concerned Scientists (UCS) \cite{UCS_DS}, number of NGSO satellites were launched into space has dramatically increased comparing to the traditional geostationary orbit (GSO) satellites. Further, this database has revealed that the total operational satellites currently in orbit around Earth are 3,000 satellites with approximately 90\% of them are NGSO systems.

These space-based Internet systems aim at providing ubiquitous, high-speed, and low-latency broadband connectivity for the fixed and mobile users. This will not only boost building a more connected world and bridge the digital divide across the globe \cite{Latio2021} but also will enable satellite systems to compete with the terrestrial networks for provisioning long-distance backhaul and directly serving users \cite{Turk2019}. In addition, the space-based Internet systems can significantly enhance the communication efficiency to markets already using satellite connectivity such as maritime, aeronautical, and big media services. Moreover, the unique features of the space-based Internet systems in providing global coverage with high speed and low-latency Internet access can create more opportunities for delay-sensitive applications to be provided through NGSO satellite systems. Thus, the next phase of developments in satellite communication systems will substantially change the way space missions are designed and operated in the near future.

\subsection{Motivation}
Motivated by the above discussions, an intriguing use case for the NGSO space-based Internet systems is to provide broadband connectivity to the nano-satellites (nanosats), i.e. space missions. In this setting, the nanosats of various downstream applications can be constantly connected to the network without depending on a private or shared distributed network of ground stations. This methodology can be also replicated for the space-based Internet providers to enable a larger degree of connectivity in space network topologies. This is certainly a game changer for the design and operation of future downstream satellite missions, where it requires the communication links to be pointing towards the sky instead of the Earth. Even though much progress has been achieved in inter-satellite links (ISLs) for low earth orbit (LEO)  systems, inter-orbit links (IOLs) have not been yet investigated for space-based Internet providers and their use has been mainly reserved for specialized GSO data relay systems as in the European Data Relay Satellite System (EDRS) \cite{EDRS}. This innovative connectivity concept can lead to more inexpensive and sustainable space systems by reducing the number of required ground stations, while achieving reliable space communications. Although, the alternatives to building vast ground station infrastructure, e.g. Amazon AWS, are specifically made for feeder links whereas NGSO constellations have multiple use cases. Besides, from a network layer perspective, nanosat broadband interconnections could be seen as a network slice of the NGSO constellations, which is already deployed for other application scenarios, e.g. residential broadband, maritime and aeronautical communications, etc.

\begin{figure}[t!]\centering
	\includegraphics[width=0.5\textwidth]{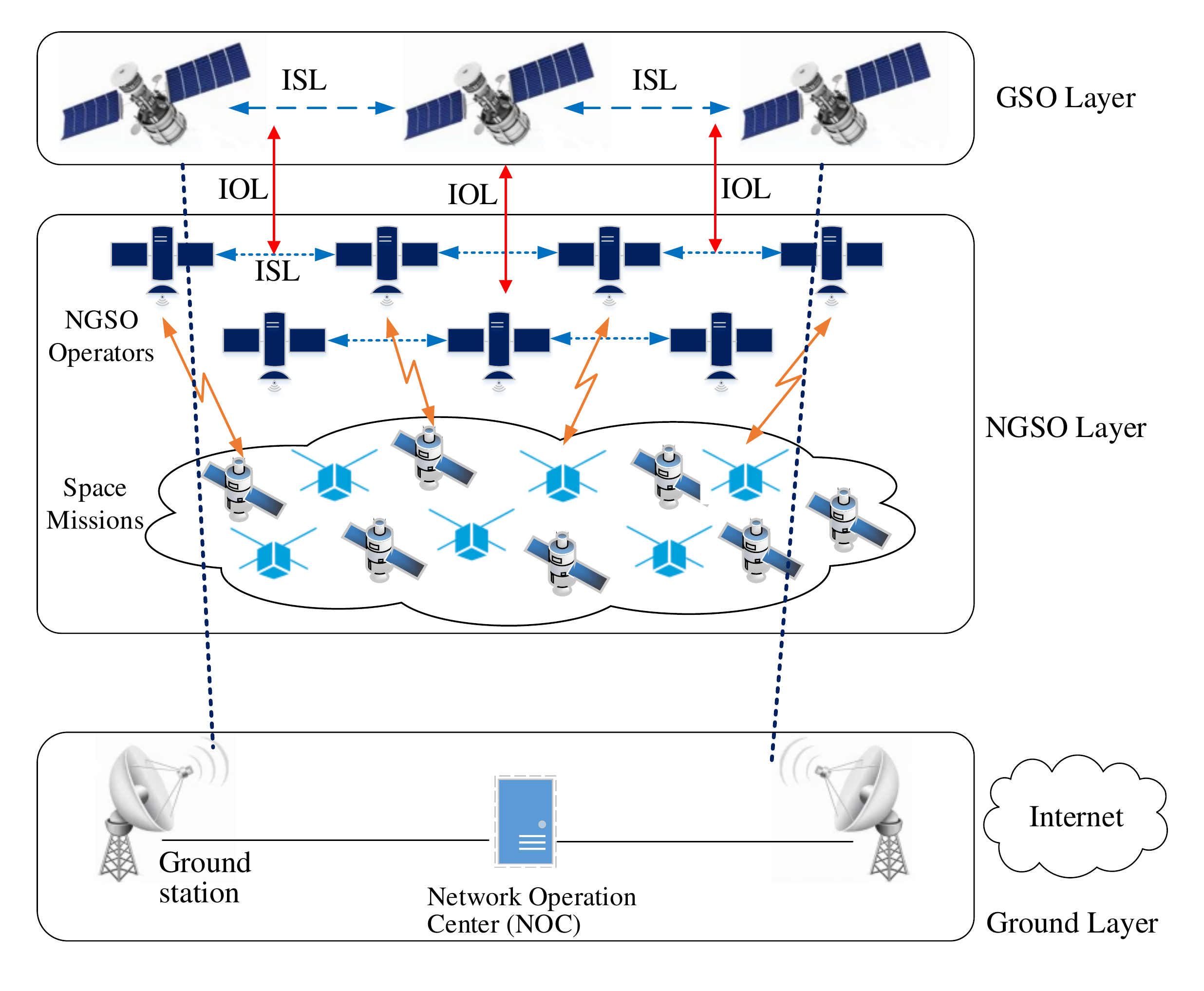}\vspace{-2mm}
	\caption{General schematic diagram of a multi-layer space information network.}
	\label{fig:SIN_structure}
\end{figure}

Developing this synergetic interaction between versatile satellite nodes across different orbits has the potential to promote connectivity and interoperability of the forthcoming satellite communications and nanosat missions. Specifically, connecting various space platforms including multi-layer multi-orbit constellation of satellites via ISLs and IOLs to construct a space information network (SIN), see Fig. \ref{fig:SIN_structure} for an illustration, can support real-time communications, massive data transmission, and systematized information services.
Through this, utilizing the space-based Internet providers for space connectivity can be seen as promising technique for nurturing the development of SIN infrastructures. Furthermore, the SINs with the improved connectivity can enhance communication and cooperation between satellites for traffic routing, throughput maximization, latency minimization, and seamless coverage.  However, interconnecting multiple satellite missions and constellations in an integrated fashion across multiple orbits will lead to a more densely connected space information system for both radio access and backhauling.

\subsection{Contributions}
In this work, we focus on studying the feasibility of using the NGSO space-based Internet systems for connecting the space missions in lower orbits. In this setting, the expected connectivity improvement will be achieved at the cost of higher complexity that is essential for load balancing between satellite links and for finding paths with the shortest end-to-end propagation delay, as well as tackling the dynamicity of the nodes (e.g. high relative speeds, frequent handovers), which is an unexplored area in the literature. 
Therefore, our objective is to investigate the most critical challenges to realize this innovative concept, in particular (a) space-based Internet access provisioning for nanosat terminals, (b) scalable space backhauling using multi-layer SINs.

The main contributions of this article can be summarized as follows:
\begin{itemize}
	\item A brief technical description is provided on the existing space-based Internet providers including their system architecture and the relevant characteristics.
	\item Identification of the most important space missions that can take advantage of a space-based Internet system and study the limitations of the new operational communication scenarios.
	\item Explore radio access schemes that can be applied in the context of SINs for connecting the beneficiary nanosats including not only the most challenging scenarios but also some potential key solutions.
	\item Examine the network architectures and routing requirements for supporting multi-tenants, multi-systems in different orbits and altitudes while being seamlessly integrated in the current Internet architecture.
\end{itemize}
The reminder of this paper is structured as follows. Section \ref{se:Sat_architecture} reviews the architectures of satellite networks, space-based Internet systems, and space missions. In Section \ref{sec:concept}, the proposed connectivity concept is elaborated from radio access and networking standpoints. Then, two case studies on both radio access design and SDN-based SIN with some simulation results are also provided in Section \ref{sec:concept}. Finally, Section \ref{sec:conclusions} summarizes the concluding remarks and gives some future research directions.

\section{Satellite System Architecture}\label{se:Sat_architecture}
Currently satellite communication systems are mainly designed separately, i.e. without considering the existence of systems simultaneously operating in different orbits. In the near future, a more integrated planning and design approaches will likely emerge to exploit the advantages of both GSO and NGSO orbits. 
Specifically, GSO and NGSO systems have their own intrinsic strengths; on one hand, GSO systems can offer broad contiguous service regions, fixed ground antenna orientation, longer satellite lifetimes, and simple interference management \cite{te_hayder2020}. On the other hand, NGSO systems offer lower latency, significantly reduced path losses that enable miniaturization of ground terminals, have the ability to cover wide areas, and support the maximum total system capacity within a certain spectrum allocation \cite{Xia2019}. Therefore, the mutual interoperability of both systems will likely converge in a shorter term allowing new satellite system architectures. This kind of satellite network structures can offer improved total spectrum utilization and increased resilience to the inter-orbit communications. 

Furthermore, the emerging NGSO satellites and mega-constellations are undergoing a significant densification in comparison to the existing GSO systems, which complicates the coordination mechanisms and overall system architecture \cite{Chen2021}. 
Establishing SIN architectures is more economically efficient and suitable for the heterogeneous integrated satellite communication systems because it eliminates the use of the excessive number of ground stations. This architecture is particularly favorable for the areas where acquiring gateway sites is difficult \cite{Hassan2020}. 
Additionally, SINs will allow a satellite system to function strategically by transmitting TT\&C data between nanosat terminals and satellite control centers on the ground. Multi-orbit SINs can provide coordination and awareness of the operational characteristics about each counterpart system, and thus, accomplish a successful coexistence between different satellite constellations without imposing detrimental interference to their concurrent transmissions \cite{Akyildiz2020}. This multi-layer SIN structure will be capable of supporting the full range of both current and future satellite service types. However, the overall network topology and space segment characteristics are the most paramount aspects in this development, which will be discussed next.

\subsection{Review of the Existing SIN Topology:}
Unlink terrestrial networks, SIN consists of a various, independent, and complex components that are designed for different purposes. The high complexity and variety of satellites along with their diverse portfolio of constellations and the high-speed mobility of NGSO with respect to the Earth’s surface inflict exceptional technical challenges on the system design and communication environment. In this direction, EDRS project \cite{EDRS} of the European Space Agency (ESA) is dedicated to the development and implementation of data relay satellites that are placed in the GSO layer to relay information to and from NGSO satellites, spacecraft, and fixed ground stations that otherwise are not able to permanently transmit/receive data. Similarly, NASA has also invested in the same concept by developing the so-called space mobile network (SMN) to be an analogous architectural framework for near earth space applications \cite{Israel2016}. 

In parallel, some works in the literature consider connecting lower orbit satellites with other higher orbit ones for routing data packets and reducing the dependency on the ground stations. For instance, the concept of system of systems was introduced in \cite{Walker2010} to study the availability and capacity of a simplified scenario consists of a few multi-orbit satellites. In \cite{Guo2020}, a futuristic architecture has been proposed based on fog environment via considering the underutilized moving satellites as mobile fog nodes to provide computing, storage and communication services for users in satellite coverage areas. 
Additionally, some other research works focus on connecting the small satellites in the same orbit via a single-layer SIN. For example, the authors of \cite{Tian2020} have proposed a self-organizing small satellite network architecture that is able to provide ubiquitous connectivity to terrestrial low-power IoT devices, where a fully distributed algorithm is developed for fast autonomous ISL establishment and channel selection that avoids interference within different ISL connections. Reference \cite{Tengyue2019} has proposed an LEO/MEO double-layer satellite network structure to utilize the advantages of both LEO and MEO satellites. This structure studies the coverage performance of satellite constellation and network transmission delay.

\subsection{Satellite System Characteristics and Classification:}
Satellite systems can be classified based on the provided services into two categories: space-based Internet providers and space missions.

\subsubsection{Space-based Internet Providers}
The space-based Internet services have been provided by multiple companies such as Hughes, Eutelsat, Viasat, and Gilat since the 1970s to regions with underdeveloped infrastructure \cite{Takei2003}. However, most of the existing systems utilize GSO satellites that are 36,000 km above the Earth, resulting in slow and expensive Internet connections. Consequently, the use of GSO-based Internet systems has been limited to a certain type of applications that are not highly susceptible to latency. In contrast, the emerging NGSO mega constellations will operate from lower altitudes, between 160 km to 2,000 km above the Earth, which lowers signal propagation loss and reduces the hardware complexity of user equipment. Several private sector companies are on their way to providing space-based Internet services in the upcoming few years, such as SpaceX, OneWeb, and SES. They have obtained licensing, launched many satellites and successfully performed initial tests. Internet giants are also foreseeing market opportunities to extend their services via NGSO constellations. For example, Amazon introduced the Kuiper project to offer high-speed broadband connectivity to people globally. Likewise, Google has invested in Starlink and supported the Loon project.

\begin{figure}[t!]
	\centering
	\includegraphics[width = 0.48\textwidth]{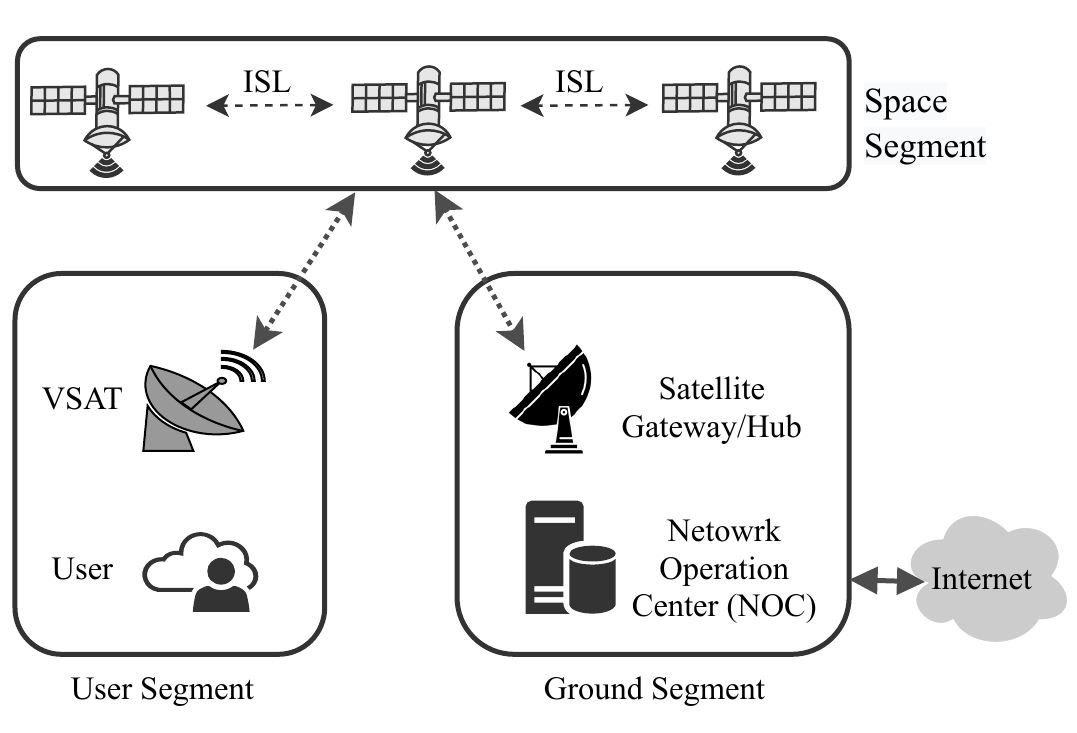}
	\caption{Schematic diagram for a space-based Internet system.}
	\label{fig:space_internet}
\end{figure}

Generally, a space-based Internet system generally consists of three main components: space segment, ground segment, and user segment (see Fig. \ref{fig:space_internet}). The space segment can be a satellite or a constellation of satellites, while the ground segment involves a number of ground stations/gateways that relay Internet data to and from the space segment, and the user segment includes a small antenna at the user location, often a very small aperture terminal (VSAT) antenna with a transceiver.
Additional critical entities within this structure are (i) network operation center (NOC) and (ii) satellite control center (SCC). The centralized NOC manages the terminal population and coordinates the information exchange with external networks, while SSC monitors and configures the network through tracking and command links.

\subsubsection{Space Missions}
The space industry is experiencing a profound change due to the miniaturization of electronic equipment to manufacture satellites leading to the emergence of new low-cost small satellites. The miniaturization of satellites is making space more affordable and accessible than ever, which will enable any country, university, startup or even school to reach space in an affordable way within a reasonable time period. Thus, these developments have unlocked the missions that satellite can carry and execute for different applications. In particular, the most relevant space missions in this context include but not limited to
\begin{itemize}
	\item Earth and Space Observation: This is one of the widespread uses of satellite constellations in different orbits including capturing high-resolution images of Earth and outer space, remote sensing in various frequencies, RF monitoring, global navigation satellite system (GNSS) reflectometry, etc.
	
	\item Asset Tracking: Satellite payload in asset tracking projects consists of a device equipped with communication components to collect information sent from objects on ground and to transmit it back to ground stations. 
	
	\item Meteorology: Nanosats are able to play an important role in storm detection and in the development of climate and weather models that enhance weather forecasts. For instance, NASA RainCube project has started the testing phase for the location, tracking and analysis of rain and snowstorms over the entire Erith.
	
	\item Agriculture: Crop monitoring is another potential use of nanosats, where a better control of harvests, the improvement of the quality of agricultural products, the finding of diseases in crops, and analysis of the ramifications derived from the periods of drought can be facilitated by using nanosats.
	 
	\item Educational Activities: The development of scientific experiments outside the Earth has become another common application of small satellites, which are unprecedented opportunities brought up by nanosats with their myriad possibilities.

	\item Government Space Programs:  The goals of these government programs varies from national security to emergency response. Some other useful applications can be for protecting the environment through detection of forest fires, studying the progress of melting ice, fighting against ocean pollution, detection of oil spills, monitoring of marine life, controlling of desertification, etc.
\end{itemize}

With these diverse applications and rapid developments in mind, there is definitely an exciting future ahead for small satellite missions in many fields but the advancement of future cooperative distributed space systems will probably require a high degree of operational autonomy. Thus, our proposed connectivity concept can be a meaningful and affordable solution for constructing independent and reliable space mission communications. 

\section{Multi-Layer Satellite Communication Systems}\label{sec:concept}
\begin{figure}[t!]
	\centering
	\includegraphics[width = 0.5\textwidth]{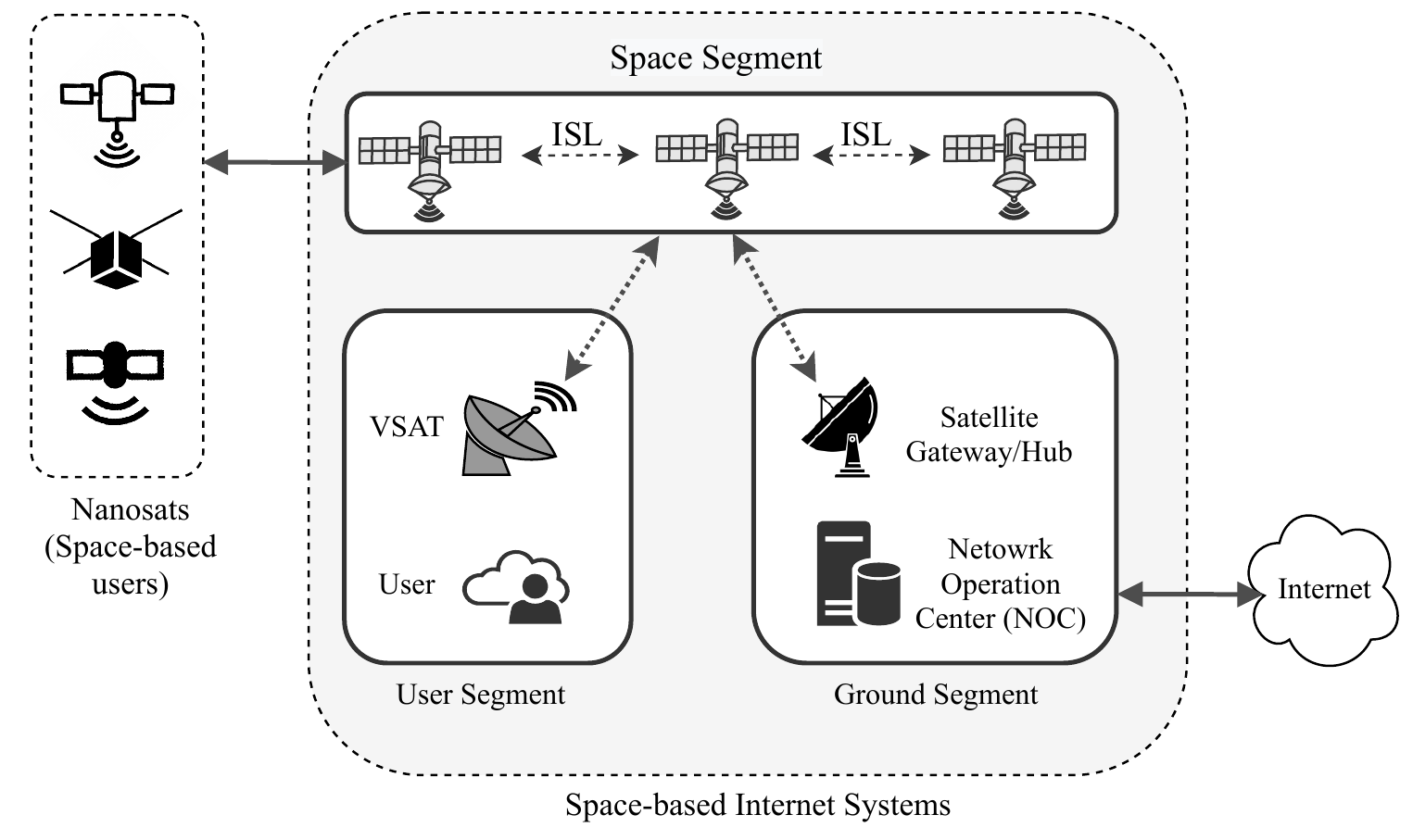}
	\caption{Nanosat terminals connected to a space-based Internet system.}
	\label{fig:system_model} \vspace{-4mm}
\end{figure}

The concept of treating nanosats as a new type of broadband users to the space-based Internet providers (as depicted in Fig. \ref{fig:system_model}) will enable combining multiple space assets to allow a more agile and efficient use of system resources. This development raises up excessive opportunities for sustainable and resilient multi-orbit satellite constellations, which can be operating as a global information network in lieu of simple bent-pipe relays. 
However, realizing this concept faces many challenges from different aspects. Specifically, granting access for such a massive number of diversified users to the space-based Internet systems while taking into account the relative motion among different entities, variable QoS requirements, differential delays, and Doppler effects are nontrivial tasks. Besides, networking and data packet routing of such a highly dynamic network considering the distinctive requirements of each information flow (e.g. TT\&C versus observation data) need in-depth analysis of the communication protocols, ISL/IOL characteristics whether RF or optical, transmit power, bandwidth along with other parameters.

Accordingly, it is crucial to study the architecture of the multi-layer SIN with focusing on the most critical challenges to realize this innovative concept, namely (a) space-based Internet access provisioning for nanosat terminals, (b) scalable space backhauling using multi-layered ISL/IOL networks. In this section, we will navigate the design issues and limitations relevant to radio access schemes and networking/routing mechanisms in the context of SINs. Further, specific design approaches and promising research directions about radio access tradeoffs and backhaul/networking requirements to realize the proposed concept will be also provided.

\subsection{Radio Access Design}\label{sec:radio_access}
One of the most important enablers for the multi-orbit multi-layer SINs is the efficiency of radio access schemes. Many access solutions for heterogeneous terminals with stationary and non-stationary channel characteristics have been developed in the framework of terrestrial networks. Herein, radio access design for nanosat terminals alongside other ground and airborne users is more complicated and challenging compared to the terrestrial case due to the different relative motion of those users with respect to satellite nodes, uneven transmit powers, link availability, and variable QoS profiles. Thereby, to serve a large number of heterogeneous users simultaneously and provide ubiquitous and flexible connectivity solutions, it is critical to devise innovative and efficient techniques that provide fair radio access and scheduling to the users in order to avoid collisions, interference, and imbalanced capacity distribution \cite{Hayder2021a}.

Prior works on access techniques have considered the large number of satellites to be deployed as a satellite sensor network, and then applied the concept of terrestrial wireless sensor networks to satellite nodes and space missions. For instance, the work in  \cite{Radhakrishnan2016} has conducted a survey on the classical multiple access protocols highlighting their benefits and pitfalls from efficiency and scalability perspectives. Authors in  \cite{Radhakrishnan2016} have also proposed two access schemes for a distributed network of nanosats, namely, (1) a modified carrier sense multiple access (CSMA) scheme that establishes communication only when it is required, and (2) hybrid TDMA and code-division multiple access (CDMA) protocol where multiple satellites from different clusters utilize same time slot using different codes. However, these developed schemes have considered only nanosats belonging to the same orbit (i.e., LEO) with typical range of 10 km to 25 km and without considering the Doppler effect by assuming a fixed distance between the satellite nodes. Reference \cite{Katona2016} has studied the multiplexing schemes for a simple use case considering one GEO satellite relays data to 15 LEO satellites and proposed two multiple radio access: (a) multi-frequency TDMA (MF-TDMA) scheme used on the high-rate links and on the low-rate telemetry links and (b) TDMA for low rate tele-command link.

The existing works on radio access design have considered only ground terminals, which essentially employ random access schemes due to the large numbers of satellite nodes and ground terminals and the fact that the traffic patterns are not known in advance. Besides, the time division multiple access (TDMA) method has been also considered and standardized in DVB-RCS (Digital Video Broadcasting-Return Channel via Satellite) specifications. Moreover, the studied system models are mostly based on small-scale scenarios and the satellite nodes act as relays. Additionally, the limitations of nanosats due to space, weight and budget, may incorporate relatively wide-beam antennas, which leads to high levels of interference that has to be properly modelled and managed. Nevertheless, accessing the space-based Internet systems that serve a large number of heterogeneous terminals from space and ground simultaneously with considering the aforementioned limitation has not been investigated yet in the open literature.

\vskip 3mm
\noindent
\textbf{Potential Solutions}\\
\vskip -3mm

Developing innovative radio access schemes for the multi-layer SINs requires major research efforts due to the spatiotemporal  heterogeneity of satellites across different orbits. In the direction of constructing the proposed concept,  we will discuss next the most promising approaches for provisioning Internet access to nanosats from the space-based Internet systems.

The multi-user MIMO technique can be utilized here to adapt to the required massive connectivity and seamless accessibility, which  is one of the most successful techniques incorporated in the terrestrial communication systems \cite{Amarasuriya2016a}. The existing beamforming methods in ground-space integrated networks are performed at the user side but recently few satellite operators have shifted this process to the satellite side as in the recent advances employed by O3b mPower satellite constellation of SES and Starlink mega-constellation of SpaceX, where a dynamic digital beamforming is conducted at the satellites. Thus, we envision this as an opportunity to facilitate the provision of Internet access to nanosats from the space-based Internet systems.

The existing beamforming algorithms depending on whether the system has knowledge of the terminal position and orientation or not. In the latter case, the beamforming will rely on the knowledge/estimation of the channel state information (CSI) and its update rate \cite{You2020}, whereas in the former case, the satellite position relative to the antenna can be readily obtained by combining ephemeris information with geometrical modelling \cite{Fu2021}. The benefit of beamforming in this context will allow delivering better signal quality to nanosats with real-time reduction in co-channel interference. Furthermore, every nanosat dictates the different type of requirement in the considered payload where three possible approaches can be highlighted; payload based on fixed beam antennas, flexible payload based on passive antennas with selectable apertures and steerable antennas, and flexible payload with reconfigurable active antennas. Therefore, these payload differences of nanosats have to be taken into account when designing beamforming techniques.

Multiplexing and diversity have the capability of offering superior performance in multiple access procedures \cite{Kibria2020}. Specifically, access schemes such as CDMA, TDMA, space division multiple access (SDMA), and probably some hybrid schemes, can be seen as feasible solutions.
Furthermore, non-orthogonal multiple access (NOMA)  is fundamentally different from the aforementioned multiple access schemes, which emerges as a solution to improve the spectral efficiency while allowing some degree of multiple access interference at receivers. NOMA can be incorporated in the multi-beam satellite architecture to design efficient transmission strategies that aim at increasing access network flexibility and capacity \cite{Lin2019}. On one hand, NOMA is more relevant in such multi-layered systems, where the wide patterns of spatial distributions of space and ground users along with the various received power levels that will cause the Near-Far Effect required for efficient NOMA. On the other hand, uplink is expected to carry more traffic than the downlink when we target nanosats due to the data-collection nature of space missions (e.g. earth imaging, meteorological applications). Thus, multiuser detection approaches must be considered in this setting, such as successive interference cancellation (SIC) techniques  or joint processing of signal copies received by multiple NGSO satellites \cite{Zhao2021}.

In short, access scheme should be tailored to the characteristics and requirements of the nanosat terminals and offered capacity by the space-based Internet providers. A not necessarily complete list of possible scenarios to be further investigated entails the following:
\begin{itemize}
	\item Nanosats suffer from less visibility time compared to ground terminals due to satellite altitude, Internet provider constellation altitude, and beam angle, and thus, applying multiplexing access protocols makes sense in this scenario. However, there is a limitation in serving a large number of users due to the constrained available resource, e.g. orthogonal codes used in CDMA are adequate to a certain number of users.
	
	\item In general, ground terminals are fixed and they often require higher throughputs than space-based terminals. However, this is not always the case for nanosats, e.g. TT\&C carriers need low-rate high-reliable links while sensor data requires high-rate downloads. This inconsistency in service demand is a crucial factor in selecting and designing proper access protocols.
	
	\item Nanosats have a better propagation environment than on-ground users owing to the reduced atmospheric attenuation and less ionosphere perturbations, which may cause harmful coexistence interference and/or receiver saturation issues. SIC multiuser detector for uplink along with NOMA concept for downlink will be employed depending on the use cases and link analysis.
\end{itemize}
More aspects and scenarios need further investigations in this direction for fostering the multi-layer satellite communication systems via devising practical ways to design affordable-complexity payload/system solutions.

\vskip 3mm
\noindent
\textbf{Case Study on Link Connectivity}\\ 
\vskip -3mm

Investigating nanosat access constraints in terms of visibility and connectivity to the space-based Internet systems is inevitable. To this end, snapshot division is a useful method for dynamic topology access design, which divides the time-varying topology into a sequence of static snapshots \cite{Madni2020}. As a preliminary analysis, we have demonstrated in the Systems Tool Kit (STK) platform a scenario with various satellites distributed across over different orbits. A space-based user (e.g. Spire satellite) with 500 km altitude and 85$^{\circ}$ inclination \cite{Spire} is assumed to directly access a satellite from the space-based Internet providers (e.g. O3b, OneWeb, Starlink). We have selected only one satellite from each constellation to analyse the visibility and periodicity with considering the distance variation as a function of time. 

\begin{figure}[t]\centering
	\includegraphics[width=0.48\textwidth]{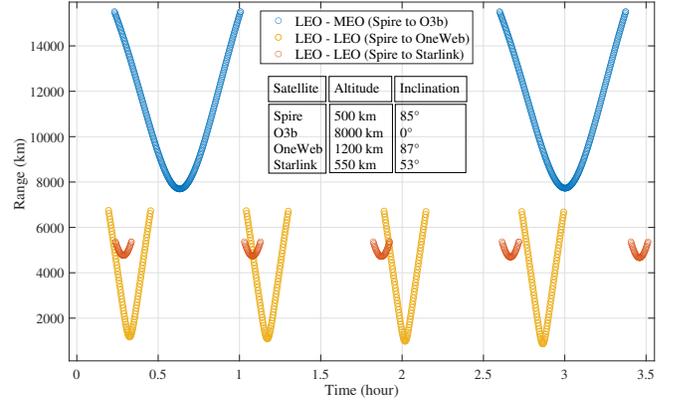}
	\caption{Range calculation and visibility analysis as a function of time between a space-based user (Spire satellite) and various space-based Internet providers (O3b, OneWeb, and Starlink).}
	\label{fig:visibility_analysis} 
\end{figure}

It can be clearly seen from the results shown in Fig. \ref{fig:visibility_analysis}, the access time duration is around 45 minutes, 15 minutes and 5 minutes for O3b, OneWeb and Starlink, respectively. The shorter coverage in LEO-LEO links is due to the high-speed movement of the satellites with respect to each other, especially in the cases when they are moving in opposite directions. In this experiment, useful information about outage and visibility can be readily extracted; obviously, the assumed space-based user is having an unceasing Internet connection from the specific space-based Internet provider for about 85\% of the experiment time. Finally yet importantly, the range variation shown in this example is larger for MEO-LEO links rather than LEO-LEO, which will have a direct impact on the round trip time (RTT) of the communication and the link budget results.

%===============================================================================
\subsection{Inter-satellite Networking and Routing}
Multi-layer SINs to a large extent is different from the traditional terrestrial networks as the former is generally composed of a multitude of satellites and constellations distributed over different orbits to provide differentiated services with various requirements, ranging from high reliability and low-rate for TT\&C subsystems to high throughput data collection for nanosats \cite{Xue2020}. 
Current satellite technologies can also allow on-board regeneration and Layer 3 routing that convert satellites to active network elements rather than simple bent pipe relays. Moreover, the dense distribution of nanosats, transmission delays, QoS priorities, uneven distribution of data flows, and the dynamic change of the network's topological structure have brought about several challenges. Specifically, the rapid increase in satellite deployments requires developing more sophisticated traffic distribution schemes to manage the growing number of satellite nodes and users to achieve network congestion control, resource utility maximization, energy efficiency, and resilience structures.  

Designing efficient routing protocols starts from evaluating the infrastructure parameters such as topology variation, bandwidth, link delay, in addition to traffic generation profiles of the heterogeneous user services/classes and computational and storage capabilities of the nodes. In the context of satellite communications, researchers have already developed several routing algorithms under the satellite network constraints. Traditional proactive and reactive routing schemes have been used in distributed and centralized systems depending on the network topology and mission requirements. These approaches require each satellite to store the entire network topology along with the routing tables \cite{Radhakrishnan2016}, but in a complex network, like our multi-layer SIN, it is difficult to maintain routing tables and consumes more power and bandwidth. 
Therefore, it is crucial to reflect the unique features of multi-layer SIN through employing space-based Internet systems, and then, designing efficient /backhauling mechanisms.

\vskip 3mm
\noindent
\textbf{Potential Solutions}\\
\vskip -3mm

It has been extensively concluded that an effective solution is given by the well-known paradigm software-defined networking (SDN) \cite{Xu2020}. SDN paradigm enables dynamic, programmatically efficient network configuration in order to improve network performance, management, and monitoring. Therefore, SDN has a tremendous potential to succeed in SINs owing to its capability to implement a reactive scheme for end-to-end traffic engineering (TE) development across the terrestrial and satellite segments. 
In the literature, prior works in \cite{Yang2016} and \cite{Sheng2017} have proposed to distribute an SDN controller on the ground, while some other works have considered the placement of the SDN controller on GEO satellites \cite{Li2017}. As intermediate solution, an SDN-based infrastructure for multi-layered space terrestrial integrated networks is introduced in \cite{Shi2019} to distribute the SDN controller entities among GEO satellites, terrestrial infrastructure, and high-altitude platforms, which is still seen as a terrestrial-dependent SDN network. Furthermore, it has been emphasised in \cite{Akyildiz2019} that there is a lack of SDN-based architecture solution specifically designed for nanosats, where all the prior works mainly focus on the traditional LEO, MEO and GEO satellites. Authors in \cite{Akyildiz2019}  have presented a detailed SDN structure adapted to the Internet of space things and nanosats but their implementation is more applicable for monitoring and Internet provisioning for remote areas, which make the developed platform terrestrial-dependent.

Employing the SDN paradigm to implement controllers on GEO satellites seems to be more suitable to deal with the constraints of the multi-layer SINs. Specifically, a small number of GSO satellites can manage to have a constant view of satellites in other orbits; however, controllers in the NGSO orbits would be also beneficial for resilience and low-latency interaction with the active network nodes in these orbits. Accordingly, the development of SDN-based SIN structure requires to deal with the accompanying challenges and to dive deep in various system solutions such as:
\vspace{-1mm}
\begin{itemize}
	\item In the study of a QoS oriented platform, the strategic selection of a path passing through several ISLs and IOLs while maintaining the end-to-end QoS requirements is quite challenging. Particularly, minimizing the controlling traffic, reducing handover procedures, and optimizing on-board flow tables’ storage will be key factors to fulfill user requirements in optimal means.
	
	\item The concept of network slicing is envisioned as a promising design approach within the multi-layer SIN structure owing to its ability of enabling optimal support for wide-reaching heterogeneous services that share the same radio access network. The coexistence of virtualized services with different QoS requirements, will make the use of SDN relevant to such dynamic network like the multi-layer SINs. In this scenario, the SDN controller is expected to have a complete view of the network. In particular, the connections among network nodes are not enough anymore, thus the controller should be able to be aware of some additional network features such as forwarding capabilities for nodes and bandwidth, delay and time-availability for links in order to optimize the overall allocation of the virtualized services.    
	
	\item The SDN concept foresees as a distributed approach where different entities of the SDN controller can manage different parts of the network. In this setting, deploying a multi-SDN controllers’ platform is quite challenging due to the unprecedented scale and complexity of such a network. Besides, hundreds of dynamic ISLs and IOLs will be used for interconnection that are constantly moving up and down expeditiously and changing in terms of supported rate, latency and topology. The high complexity of this SIN requires network-level optimization algorithms for distributing SDN controller entities aiming at minimizing the controlling traffic.
\end{itemize}
Thus, developing effective routing algorithms and designing in-space backhaul for the multi-layer SINs require to put forth more research efforts to consider diverse data flows and the heterogeneity of the served/embedded slices as well as the huge size and dynamicity of the systems. Additionally, several well-developed network emulators and simulation tools such as Mininet and AGI STK alongside with the open-source SDN controller can be utilized for performance evaluation. Specifically,  a simulation platform that supports various mobility traces and protocols for the multi-layer SIN can be developed by using the distinguished network emulator Mininet that facilitates defining system parameters while instantiating the link/network devices such as bandwidth, delay, packet loss, computation capability, which will help constructing a data-plane network that combines the multiple heterogeneous elements of the multi-layer SIN.

\vskip 3mm
\noindent
\textbf{Case Study on SDN-based Multi-layer SIN:}\\ 
\vskip -1mm

\begin{figure}[!t]\centering
	\includegraphics[width=0.5\textwidth]{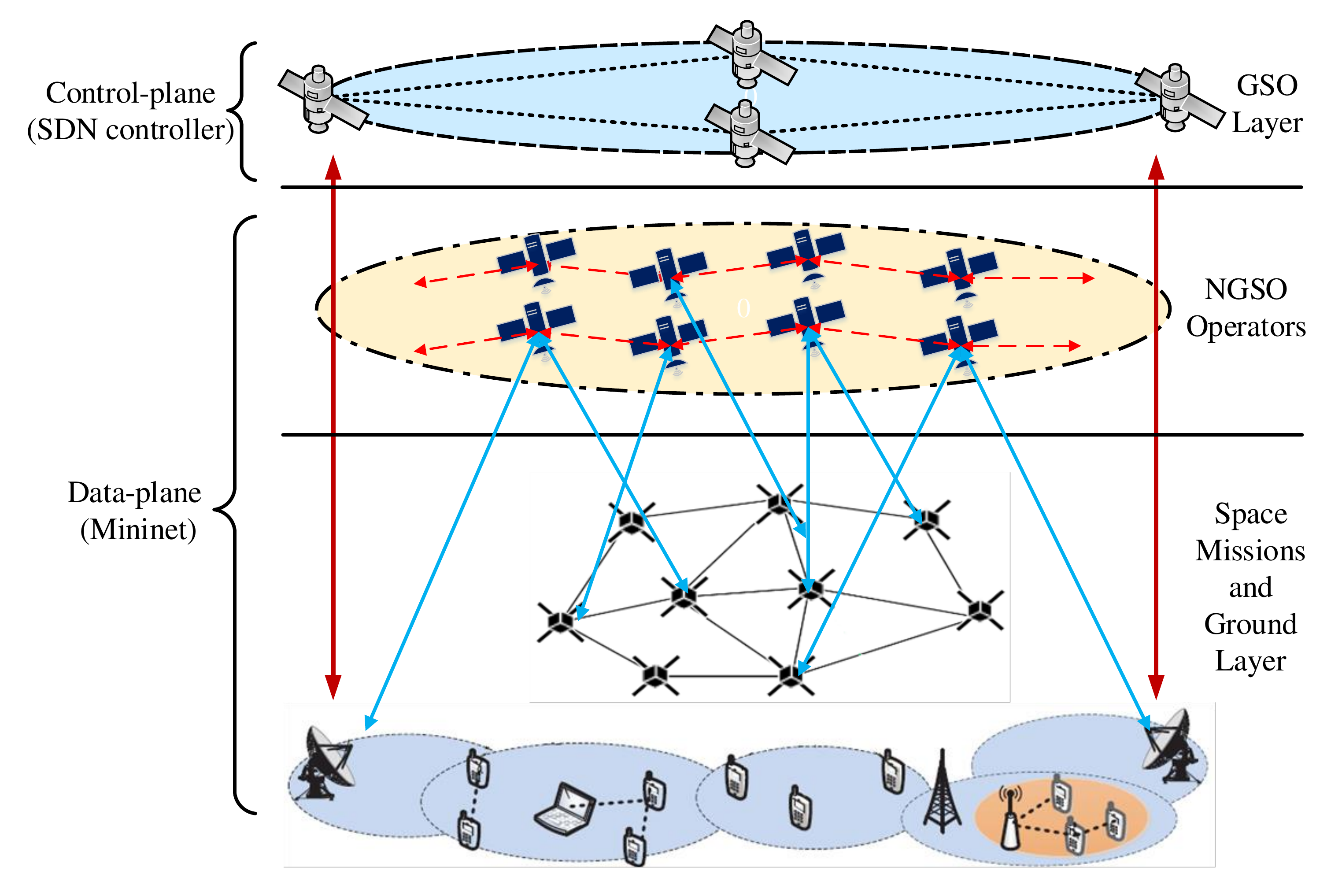}
	\caption{SDN-based architecture of multi-layer SIN.}
	\label{fig:SND_based_SIN}\vspace{3mm}
\end{figure}

The global view offered by GEO layer to the network below (from NGSO layer until the terrestrial layer) can be exploited for establishing and running the SDN-controllers as illustrated in Fig. \ref{fig:SND_based_SIN}. However, this is not the only scenario for efficient SDN-controllers distribution as the controllers can be placed in lower layers to ensure resilience on low latency when only ISLs are involved. In this scope, an SDN application can be running in the controller to compute the most efficient path for a single service. The computational complexity will be influenced by QoS requirements of the service such as latency and demanded bandwidth and by the physical network status such as the actual traffic across the network, bandwidth availability, storage capability for the intermediate nodes. 

In this setting, the SDN controllers can utilize the well-known OpenFlow protocol to facilitate SDN realization.  OpenFlow protocol \cite{OF_specification} will facilitate SDN controllers to translate the computed data path into flow rules for the switches, enabling the packet forwarding toward the following hop.
This separation between control and forwarding plane allows for more sophisticated traffic management than is feasible using access control lists (ACLs) and routing algorithms. Additionally, OpenFlow tolerates switches from different vendors, often each with their own proprietary interfaces and scripting languages, to be managed remotely using a single open protocol. For large, dynamic and heterogeneous networks, such as multi-layer SIN, the choice of a robust and distributed SDN controller will be relevant to successfully manage a large quantity of setup messages. 

\begin{figure}[!t]\centering
	\includegraphics[width=0.48\textwidth]{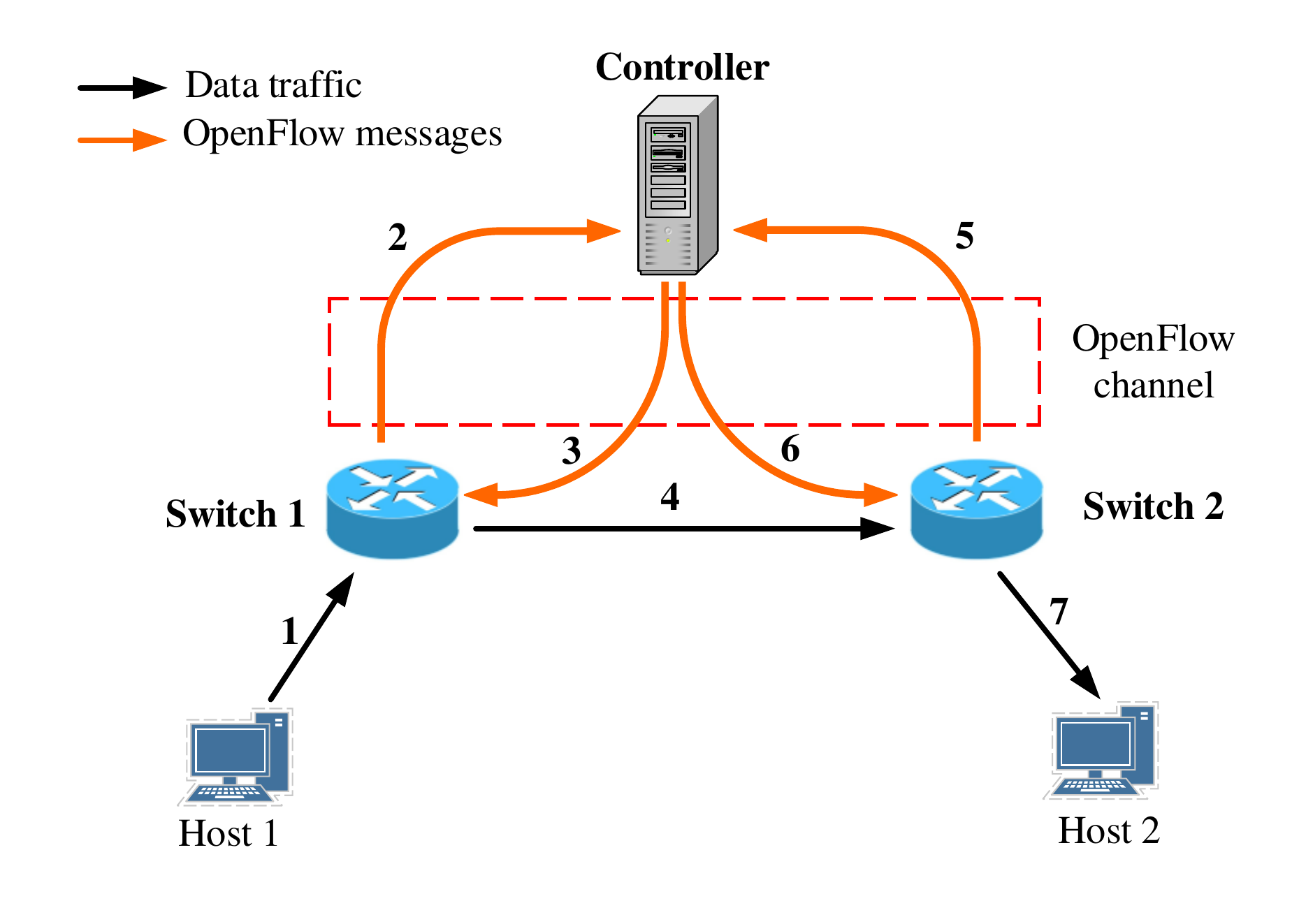} 
	\caption{Controller-satellite sequence of OpenFlow messages.}
	\label{fig:OF_mess} 
\end{figure}

In the following, we present some preliminary findings about the messages exchange between the SDN controller and network nodes (switches). Fig. \ref{fig:OF_mess} shows a simplified example for OpenFlow channel when a new link is set up and the traffic from Host 1 to Host 2 is being assigned to that link. It is worth underlining that the term "\textit{new}", in this section, is used to describe the scenario where the switch has not received any indications from the controller, on which traffic has to be forwarded to the new link, yet. Each network node is considered as OpenFlow-enabled switch. When a packet is arriving to Switch 1 and the routing application, running in the SDN controller, has decided (due to pre-defined routing policy) that it should be forwarded to the new link, the process over OpenFlow channel is started. It can be shown that at least 4 messages are exchanged per each new established hop and per each direction of the traffic. According to the sequence presented in Fig. \ref{fig:OF_mess}, the Switch 1 initially asks (message 2) the controller to install flow rules for the received packet. Flow rules constitute, in each switch, the forwarding table entries. Each entry gives the instruction to the switch on which port the incoming traffic should be forwarded. The traffic is selected based on OpenFlow "\textit{matching criteria}" (ingress port, MAC-IP source/destination addresses, TCP/UDP traffic, etc.). In the SDN controller, a pre-defined routing application computes the next hop for the received packet(s). Consequently, the controller replies with message 3 (flow rules to be installed in the switch). In this particular case, the flow rule is giving indication to the Switch 1 that, the packets coming from Host 1 with destination Host 2, must be forwarded to Switch 2. Afterwards, the traffic arrives to Switch 2 and the same process is repeated (messages 5 and 6). 
In addition, when the traffic is going in the opposite direction, from Host 2 to Host 1, the same flows of messages controller-switch runs. For the sake of simplicity, we have considered a simplified version of only 1 hop source and destination. However, the process is repeated for each new hop in the path. The number of messages, exchanged between controller and each network device, is in general higher than what we have described here, however this discussion gives the basis to understand the minimum signalling traffic over the network when considering a SDN-based implementation.   

Given these premises, every time there is a change in the network, there might be one or more affected devices. it is highly probable that the affected devices have to change their path to reach the destination. For each hop of the new path, the exchange of messages from controller to OpenFlow enabled device happens. If we consider bidirectional communications, the results have shown that at least 8 OpenFlow setup messages, per new hop, flow over the channel. This means that if the amount of affected devices is high and the data paths are long, the number of messages to be communicated through the controllers increases considerably. 

\begin{figure}[t]\centering
	\includegraphics[width=0.5\textwidth]{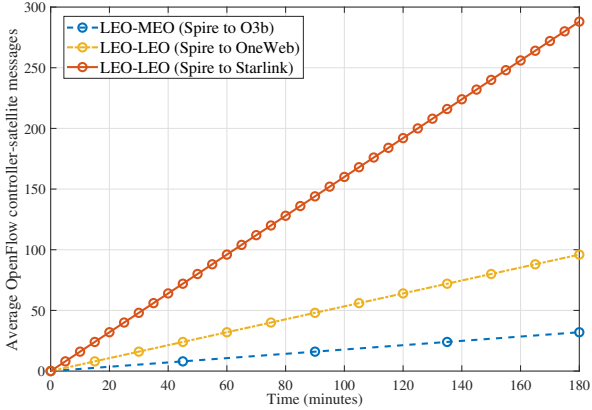}
	\caption{Average messages controller-device exchange for different scenarios of connection from a space-based user (Spire Satellite) and various space-based Internet providers (O3b, OneWeb, and Starlink).}
	\label{fig:SDN_messages_graph} 
\end{figure}

Following the three scenarios presented in the case study on link connectivity presented in Section \ref{sec:radio_access}, we have performed a simulation to analyze the load of messages which would arrive to the SDN controller for the three different scenarios. Fig. \ref{fig:SDN_messages_graph} shows the average OpenFlow messages exchanged between the controller and the nodes of the multi-layer SIN, every time there is new hop modification, versus time. For the Spire-Starlink scenario, every 5 minutes, there is a variation in the network connectivity, which corresponds to a new setup from SDN controller's side. For the Spire-OneWeb scenario, the interval of time is every 15 minutes and for Spire-O3b is 45 minutes. The simulation shows that, considering a simulation of 180 minutes, the SDN controller, in the first scenario, is required to manage a quantity of messages that is 3 times more than the Spire-OneWeb case and up to 9 times more than the one for Spire-O3b link. This is just to give an idea of the extreme capability that the SDN controller is expected to have in the scenario of multi-layer SIN. 

Thus, it is necessary to have a solid SDN controller capable of instantiating different entities when it is needed, and at the same time keeping under control, while  minimizing, the signalling traffic. Additionally, it is worth underlining that the OpenFlow messages are signalling traffic due to the fact that they do not carry information, but still they consume bandwidth. Having a controller that is able to find the best compromise between reduced signalling traffic and reliable connection, will be a key factor to successfully address  the development challenges of multi-layer SINs.

\section{Conclusions and Future Research}\label{sec:conclusions}
This article presents a disruptive approach to circumvent the connectivity issues encountered by various space downstream satellites in lower orbits via establishing a multi-layer SIN utilizes the advent space-based Internet providers to provide persistent broadband connectivity. We have also discussed the state of the art in satellite networks, NGSO space-based Internet systems, and nanosat space missions. The proposed connectivity model can unleash nanosat potentials and allow a larger degree of connectivity for space network topologies. There are however many technical challenges that need to be adequately addressed to realize such network structures, including the design of compatible radio access schemes and the definition of efficient SDN-based satellite networking and routing mechanisms, which are the most critical challenges and the main focus of this article. More specifically, radio access and networking challenges are carefully examined within the proposed multi-layer SIN with emphasizing on the stringent technical limitations and requirements. Beyond this, some potential solutions to deal with the identified challenges and to fulfill the requirements are also provided in order to move forward towards realizing this interesting setup.

The contribution of this paper paves the way for a systematic and seamless integration of the nanosat missions with the space-based Internet systems, thereby stimulating further research in this domain. Most importantly, our future studies will be focused on developing and evaluating of optimal service-oriented radio access schemes and efficient routing algorithms. Meanwhile, more futuristic visions and innovative research directions can be inspired by utilizing the proposed concept such as empowering satellite constellations with storage and processing capabilities in a virtualized manner but that would lead to even more complicated slicing scenarios. Furthermore, the edge computing technique, which was emerged for lessening the high latency issue, would be more meaningful and feasible in the context of multi-layer SINs. Particularly, heavy computational tasks, such as online optimization of the resource allocation strategy, data processing for earth observation applications, data aggregation for IoT, etc., can hardly be executed  using a single small satellite processor; alternatively, such kind of tasks can be done in a distributed fashion. In short, the proposed architecture in this article has the potentials to create new opportunities for the satellite operators and revolutionize the design and deployment of future satellite missions.

\linespread{1.2}
% ===========================================================================
% bibliography
% ===========================================================================
\bibliographystyle{IEEEtran}
\bibliography{IEEEabrv,References}
\end{document}